# INTERPRETATION OF QUANTUM THEORY AND COSMOLOGY

G. M. Prosperi and M. Baldicchi

*Dip. di Fisica dell'Università di Milano*
*I.N.F.N., sezione di Milano*

giovanni.prosperi@mi.infn.it

**Abstract**

We reconsider the problem of the interpretation of the Quantum Theory (QT) in the perspective of the entire universe and of Bphr idea that the classical language is the language of our experience and QT acquires a meaning only with a reference to it.

We distinguish a classical or macroscopic level, and a quantum or microscopic one that is perceived only through the modifications that it induces in the first. The macroscopic state of the universe is assumed to be specified by a set of variables , a classical energy momentum tensor and some conserved currents, which are supposed to have a well defined value across the entire space-time. To the energy-momentum tensor a classical metric *is related* by the Einstein equation.

The quantum state and dynamics are expressed by the usual QT formalism in terms of a density operator and the ordinary quantum operators in Heisenberg picture. For the macroscopic variables a basic distribution of probability is postulated in terms of a density and the corresponding quantum operators, so in some way their evolution is driven by the underlying QT. Such postulate essentially replaces the usual selfadjoint operators correspondence

For the Universe we adopt a variance of the $\Lambda$CDM model with $\Omega = 1$, , one single inflaton with an Higgs type potential, the initial time at $t = -\infty$. The expectation values of all fundamental fields are supposed to vanish for $t \to -\infty$ . In the framework the scalar fluctuation in the Cosmic Microwave Background are correctly explained giving appropriate calue to the parameters in the potential. As in more conventional models the absence of the tensor fluctuations remains not understood, if even a quantum metric is introduced. This seems to suggest that Gravity is a pure classical phenomenon, what could be consistently accommodated in our formalism by an appropriate even if somewhat ad hoc assumption.

## 1. Introduction

In cCassical Physics the state and the evolution of a system is described in terms of a set of quantities having always a well-defined value. On the contrary in quantum physics it is not possible to ascribe to a quantity a definite value independently of an actual observation. This is an obvious consequence of the fact that the probability to find a certain result depends quadratically on the state vector, while the evolution of this obeys a linear equation. In order we can talk of an observation, however, the result must be stated in definite terms.

For this reason Bohr [1] maintained that Quantum Theory (QT) applies to objects for which no model from our ordinary intuition would be appropriate, but the experimental setup and the results of any experiment should be expressed in classical terms; classical language being the language of our experience. An obvious immediate objection to Bohr's point of view is, however, that all experimental apparatuses are thought to be made by the same particles to understand which behaviour QT has been developed.

To overcome this difficulty von Neumann proposed a theory of measurement in which even the apparatus is treated according Quantum Theory [2]. Precisely, given a certain object, to be denoted by I , let us consider an apparatus, denoted by II , to observe a quantity $A_I$ relative to the object, represented by a certain selfadjoint operator $\hat{A}_I$ . The apparatus must be a system which interacts for a short time with the object and reacts in a different way according the eigenstate of $\hat{A}_I$ in which I is. Let be $\psi_1, \psi_2, \ldots \psi_r, \ldots$ the eigenvectors of $\hat{A}_I$ and $\alpha_1, \alpha_2, \ldots \alpha_r, \ldots$ the corresponding eugenvalues, $\Psi_0, \Psi_1, \ldots \Psi_r, \ldots$ the eigenvectors of a certain other appropriate quantity $M_{II}$ relative to II with eigenvalues $\mu_1, \mu_2, \ldots \mu_r \ldots$ . We assume that, if I is initially in the state $\varphi_s$ and II in the state $\Phi_0$ , as result of the interaction one has

$$\psi_s \Psi_0 \to \psi_s \Psi_s . \qquad (1.1)$$

Then, if we consider I in the more general state $\psi = \sum_s c_s \psi_s$ , we have

$$\psi \Psi_0 \to \sum_s c_s \psi_s \Psi_s . \qquad (1.2)$$

Consequently the probability $|c_s|^2$ to observe the value $\alpha_s$ for $A_I$, according Quantum Mechanics when applied to the system I , is identical to the probability to observe $\mu_s$ for $M_{II}$ after the interaction, when QT is applied even to II. Naturally, if we apply QT even to II , we must refer to a third system III described classically acting as apparatus for the measurement of $M_{II}$ on II . In this way a chain of apparatuses is created, every one observing the preceding one, extending to the nervous system of the human or conscious observer. The important point is that to apply QM to I and describe classically II+III is completely equivalent to apply QM to I + II and describe classically III. That is, in a sense the world is split in two parts, one described according to QM regarded as the object and another regarded as the observer that is described classically and includes all conscious beings, but the border between the two can be shifted arbitrarily. In last analysis who decides what happened should be the conscious act of an "abstract ego".

Note that, due to the entanglement which is established between the states of the nervous systems of two human observers looking to the same apparatus, Von Neumann interpretation is intersubjective and not subjective. However, two main observations are in order. The first concerns eq. (1.1) itself. Obviously the Hamiltonian that controls the evolution of a systems and in particular its interaction with other systems is not at our choice. Therefore the equation can be realized only in specific cases and so the class of the actual observables is much more restricted than that of of the selfajoint operators, as it is supposed in the ordinary formulation of QT. The second one is that, whatever philosophical attitude we can take, what we experience must be related to a specific system of orthogonal states of our brain, with the exclusion of a superposition of them.  So the meaning itself of the theory should be referred to the consideration of this systems, that does not seems particular significant from the physical point of view.

It seems necessary to  be able to break the von Neumann chain at a certain level and justify the use of the classical language beyond this stage. Many attempts have been made in this sense, but none is satisfactory in our opinion. For a sample of the debates still ongoing in 1971 see ref. [3] and for a rather complete reference before 1983  and a reproduction of a set of select papers  see [4].

Obviously, as matter of  fact in any specific application of QT the line of demarcation is set where we reach the macroscopic level. In ref. [5] our group in Milan had attempted to give a formal basis to some general ideas of P. Jordan [6], according to which in a systems made by a large number of components the interference terms between macroscopic distinguishable states should largely cancel each other. In particular we made reference to an ergodic assumption in Quantum Statistical Mechanics or to other equivalent properties that enable to derive the usual deterministic equations of classical Physics for a latge body-.

The point of view was contrasted by Wigner (see [7] and  references therein), who stressed that, being the argument  essentially based on approximations, what really we can prove is that the observation of interference terms between macroscopic distinguishable states is very difficult, but in principle they remain there, as soon as we are in the framework of the ordinary formulation of QM.

Taking into account the above criticism, in ref. [8] and particularly in ref. [9], we tried to overcome the difficulty postulating that the macroscopic variables have to be formally treated as continuously observed and so supposed always well defined. This can be done in the framework of a theory of a continuous monitoring of a system,  made in  the context of a generalized formulation of Quantum  Mechanics (GQM, [10-12]), in which observables are related to *positive operator measure* (pom) rather than  simply *projector measure* as in the ordinary one. It is this that enables to avoid the so called Zeno theorem.

The continuous approximate  monitoring of a set of observables , $A_1, A_2, \ldots A_p$, compatible or not, causes however a disturbance on the system which in Schroedinger picture can be expressed as a modification of the Liouville - von Neumann equation as (we shall use as a rules natural units and 1 GeV  as fundamental unit))

$$\frac{\partial \widehat{W}}{\partial t} = -i\left[\widehat{W}, \widehat{H}\right] - \sum_{s=1}^{p} \gamma_s \left[\hat{A}_s, \left[\hat{A}_s, \widehat{W}\right]\right], \qquad (1.3)$$

where  $\widehat{W}$ is the statistical or density operator, $\hat{A}_1, \hat{A}_2, \ldots \hat{A}_p$ are the quantum operator corresponding to $A_1, A_2, \ldots A_p$ and  $\gamma_1, \gamma_2, \ldots \gamma_p$ are real constant that have to be supposed positive if we want  $\widehat{W}$ to remain a positive operator.  In the perspective in consideration, the operators  $A_s$, were identified as the variables specifying the macroscopic state of the system. In ref. [13] an equation similar to (1.3) was considered as a new assumption independently of a reference to a continuous observation and it was shown  that, as a consequence of it, the interference terms among different eigenstates of the $\hat{A}_s$ vanish as the time increases. Note however that the additional term in (1.3) violates the energy conservation if $A_1, A_2, \ldots A_p$ are not all constant of the motion. Furthermore a significant relativistic extension of the equation does not seem possible. An attempt to overcome the above difficulties releasing the requirement of positivity for the $\gamma_s$ made in ref. [14], does not seem to bring to a sufficiently general class of classical observables.

In the present paper we shall reconsider the problem in the frame of the entire universe, by distinguishing a classical or macroscopic level, which is what it is related to our experience according to Bohr's views, and a quantum or microscopic level that is perceived only through the modifications that it induces on the macroscopic level. We shall adopt a ΛCDM cosmological model with $\Omega = 1$, but wiyh the the initial time at $t = -\infty$, as required, as we shall see, by the set of our assumptions.

Inspired by the most general framework of Classical Physics, which is the non-equilibrium Thermodynamics, we assume the macroscopic state of the universe to be characterized by a classical total energy momentum tensor $t^{\mu\nu}(t, \boldsymbol{x})$ and classical conserved currents, $j_1^\mu(t, \boldsymbol{x}), j_2^\mu(t, \boldsymbol{x}), \ldots j_r^\mu(t, \boldsymbol{x})$ (e. m. current, barion and lepton currents, if the ordinary particle standard model is adopted). They are supposed to have always a well-defined value at any time and place or better to be represented by well-defined distributions in the sense of Schwarz. To the *classical tensor* $t^{\mu\nu}(x)$ we assume to be associated a *classical metric* $g_{\mu\nu}(x)$ defined by the *Einstein equation*

$$R_{\mu\nu}(g_{\rho\sigma}) - \tfrac{1}{2} g_{\mu\nu}(x) R(g_{\rho\sigma}) = -8\pi G \, t^{\mu\nu}(x) \, , \qquad (1.4)$$

$R^{\mu\nu}(g_{\rho\sigma})$ being the Ricci curvature tensor relative to the metrix $g_{\mu\nu}(x)$, $R = g_{\mu\nu} R^{\mu\nu}$ the scalar curvature, $G$ the gravitational constant. From (1.4) by the Bianchi identity the conservation equation follows

$$D_\mu t^{\mu\nu} = \partial_\mu t^{\mu\nu} + \Gamma^\mu_{\mu\rho} t^{\rho\nu} + \Gamma^\nu_{\mu\rho} t^{\rho\nu} = 0 \, . \qquad (1.5a)$$

We postulate also

$$D_\mu j_k^\mu = \partial_\mu j_k^\mu + \Gamma^\mu_{\mu\rho} j_k^\rho = 0 \qquad (k = 1, 2, \ldots r) \, . \qquad (1.5b)$$

To make unique $g_{\mu\nu}(x)$, once that $t^{\mu\nu}(x)$ is given, we assume that in an appropriate reference it becomes only time dependent (homogeneous in space) and it tends to the a Robertson-Walker form

$$\bar{g}'_{00}(x) = -1, \quad \bar{g}'_{0j}(x) = 0, \quad \bar{g}'_{ij}(x) = a'^2(t)\, \delta_{ij} \quad \text{with} \quad a'(t) \to 0 \qquad (1.6)$$

as $t \to -\infty$.

In a similar way a classical e. m. field can be associated to the e. m. current by the Maxwell equations.

On the contrary the microscopic level is supposed to be described in terms of the usual quantum formalism and the microscopic state defined by a positive trace one statistical operator $\widehat{W}$. No modification is introduced in the quantum dynamics, Heisenberg picture is used as a rule and $\widehat{W}$ is correspondingly supposed independent of time. We assume the total action to be expressed in terms of a quantum metric $\gamma_{\mu\nu}$ and of a number of fundamental fields $\phi_1(x)$, $\phi_2(x), \cdots$ depending on the specific elementary particle model adopted. Among such fields, we assume to be one scalar *inflaton field*, which we shall denote specifically by $\varphi(x)$. Wishing to put in evidence the latter, we can write the total action of the universe as

$$S_T = S_\gamma[\gamma_{\mu\nu}] + S_\varphi[\gamma_{\mu\nu}, \varphi] + S_\phi[\gamma_{\mu\nu}, \varphi, \phi_1, \phi_2, \cdots], \qquad (1.7)$$

where

$$S_\gamma[\gamma_{\mu\nu}] = \int d^4x \sqrt{-\gamma}\, R(\gamma_{\mu\nu}), \qquad (1.8)$$

$\gamma = \det \gamma_{\mu\nu}$,

$$S_\varphi[\gamma_{\mu\nu}, \varphi] = -\int d^4x \sqrt{-\gamma}\, \left(\tfrac{1}{2}\gamma^{\rho\sigma}\partial_\rho\varphi\,\partial_\sigma\varphi + V(\varphi)\right), \qquad (1.9)$$

and we take $V(\varphi)$ to be the renormalizable potential

$$V(\varphi) = \frac{\kappa^4}{4\lambda} - \frac{1}{2}\kappa^2\varphi^2 + \frac{1}{4}\lambda\varphi^4. \qquad (1.10)$$

The interaction between $\varphi$ and any other field is supposed included in the last term in (1.7).

Finally, as quantum state initial condition, we assume that for $t \to -\infty$ the universe is in its vacuum state $\widehat{W}_{in} = |0\rangle\langle 0|$. Then the expectation values of all spinor and vector quantum fields vanish

$$\langle \hat{\phi}_s(x)\rangle_Q \equiv \text{Tr}[\hat{\phi}_s(x)\widehat{W}_{in}] = \langle 0|\hat{\phi}_s(x)|0\rangle \to 0. \qquad (1.11)$$

The situation is more complicate for the inflaton and the other scalar fields for which $\bar{\varphi} = \langle 0|\hat{\varphi}(x)|0\rangle$ can be non zero. In the context we assume again the vacuum invariant under three dimensional space translation and rotation (homogeneity and isotropy of the space) but not under time translation. Therefore $\bar{\varphi}$ must depend on the time alone and we write

$$\hat{\varphi}(x) = \bar{\varphi}(t) + \hat{\chi}(x) \quad \text{with} \quad \langle \chi(x)\rangle_Q \to 0. \qquad (1.12)$$

Correspondingly we can write

$$\hat{\gamma}_{\mu\nu}(x) = \bar{g}_{\mu\nu}(t) + \hat{h}_{\mu\nu}(x) \quad \text{with} \quad \langle h_{\mu\nu}(x)\rangle_Q \to 0 \qquad (1.13)$$

$\bar{g}_{\mu\nu}(t)$ being of the same type as in eq. (1.6)

$$\bar{g}_0(x) = -1, \quad \bar{g}_{0j}(x) = 0, \quad \bar{g}_{ij}(x) = a^2(t)\,\delta_{ij}. \qquad (1.14)$$

We shall suppose $\bar{\varphi}(t)$ and $\bar{g}_{\mu\nu}(t)$ to be solutions of the Euler equation relative to the action (1.7-10) which when all other fields are set to 0 take the form

$$\ddot{\varphi}(t) + 3\frac{\dot{a}(t)}{a(t)}\dot{\varphi}(t) + V'(\varphi(t)) = 0 \qquad (1.15)$$

and of the Friedman equation

$$\frac{\dot{a}(t)}{a(t)} = \sqrt{\frac{8\pi G}{3}\left(\frac{1}{2}\dot{\varphi}(t)^2 + V(\varphi(t))\right)}, \qquad (1.16)$$

with the initial conditions

$$\bar{\varphi}(t) \to 0 \quad \text{and} \quad a(t) \to 0$$

in order to have (1.11) satisfied for every field and shall take $\hat{h}_{\mu\nu}(x)$ as the definition of the quantum gravitational field.

Let us come to the dynamics of the macroscopic variables, $t^{\mu\nu}(x)$ and $j_k^\mu(x)$ (with $k = 1, 2, \ldots r$). It shall be expressed by a probability distribution, which we shall postulate in the next section in terms of the quantum operators $\hat{T}_{\mu\nu}(x)$, $\hat{J}_{k\mu}(x)$, $\hat{\gamma}_{\rho\sigma}(x)$ and the statistical operator $\widehat{W}$. This postulate has to be understood as a basic assumption, which in some way replaces the ordinary selfadjoint operators observables correspondence.

Since the classical variables as well the quantum operator are supposed, as we said, to be distributions, only expressions of the type are significant

$$\vartheta_f = \int d^4x \, f^{\mu\nu}(x) \, t_{\mu\nu}(x) \quad \text{and} \quad \sigma_{kh} = \int d^4x \, h^\mu(x) \, j_{k\mu}(x) \,, \tag{1.17}$$

where $f^{\mu\nu}(x) = f^{\nu\mu}(x)$ and $h^\mu(x)$ are assumed to be sufficiently regular functions (weight functions), which vanish rapidly in all space-time directions and transform as tensor or vector densities in order the quantities (1.17) be coordinate independent.

A typical choice could be

$$f^{\mu\nu}(x) = e^{\mu\nu} w(x) \quad \text{or} \quad h^\mu(x) = e^\mu \, w(x) \,, \tag{1.18}$$

Where $e^{\mu\nu}$ and $e^\mu$ project on a specific component and $w(x)$ a positive function wich vanishes out of a small neighbour of a space-time point $x_w \equiv (t_w, \boldsymbol{x}_w)$ and $\int d^4x \, w(x) = 1$; e. g.

$$w(x) = \frac{\Lambda_t \Lambda_s^3}{\pi^2} \exp\left[-\{\Lambda_t^2 (t - t_w)^2 + \Lambda_s^2 (\boldsymbol{x} - \boldsymbol{x}_w)^2\}\right] \,, \tag{1.19}$$

with $\Lambda_t$ and $\Lambda_{st}$ sufficientely large.

This understood, in sec. 2 we shall postulate a density of probability of the type

$$p(\vartheta_{f_1}, \vartheta_{f_2}, \ldots \,; \, \sigma_{1h_{11}}, \sigma_{1h_{12}}, \ldots \,; \, \sigma_{2h_{21}}, \sigma_{2h_{22}}, \ldots \,; \ldots \,; \, t_b, t_a \,; \, \widehat{W}_a) \tag{1.20}$$

for any number and choice of the weight functions which satisfy some additional conditions too that we shall discuss in the following. As indicated, such expression refers to a given time interval $(t_a, t_b)$ and to a given microscopic state, specified by the statistical operator $\widehat{W}_a$. The latter is supposed to reflect the story of the Universe before $t_a$. In particular for $t_a = -\infty$, we must have $\widehat{W}_a = \widehat{W}_{\text{in}} = |0><0|$..

The assumed expression for (1.20) has some important properties, that we shall prove in the next, but it is convenient to anticipate in order to clarify them.

First, we have the expectation values

$$\langle \vartheta_f \rangle = \langle \widehat{T}_f \rangle_Q \quad \text{and} \quad \langle \sigma_{kh} \rangle = \langle \hat{J}_{kh} \rangle_Q \,, \tag{1.21}$$

with

$$\widehat{T}_f = \int d^4x \, f^{\mu\nu}(x) \, \widehat{T}_{\mu\nu}(x) \quad \text{and} \quad \hat{J}_{kh} = \int d^4x \, h^\mu(x) \, \hat{j}_{\mu k}(x) \tag{1.22}$$

and having again denoted by $\langle \hat{A} \rangle_Q$ the Quantum expectation value of the quantity $A$

$$\langle \hat{A} \rangle_Q = \text{Tr} \, (\hat{A} \widehat{W}) \,. \tag{1.23}$$

Going back to eq. (1.6) it is then natural to identify $a'(t)$ with $a(t)$ in (1.14), $\bar{g}'_{\mu\nu}(t)$ with $\bar{g}_{\mu\nu}(t)$ and define a classical gravitational field $h'_{\mu\nu}(x)$ by $g_{\mu\nu}(x) = \bar{g}_{\mu\nu}(t) + h'_{\mu\nu}(x)$.

For the variances we have similarly

$$\langle (\vartheta_f - \langle \vartheta_f \rangle)^2 \rangle =$$
$$= \frac{\mu^4}{4} \int d^4x \, \langle (-\hat{\gamma})^{-\frac{1}{2}} \hat{\gamma}_{\mu\rho} \hat{\gamma}_{\nu\sigma} \rangle_Q \, f^{\mu\nu}(x) f^{\rho\sigma}(x) + \langle (\widehat{T}_f - \langle \widehat{T}_f \rangle_Q)^2 \rangle_Q \,, \tag{1.24 a}$$

$$\langle (\sigma_{k\,h} - \langle \sigma_{k\,h} \rangle)^2 \rangle =$$
$$= \frac{v_k^3}{4} \int d^4x \, \langle (-\hat{\gamma})^{-\frac{1}{2}} \hat{\gamma}_{\mu\rho} \rangle_Q \, h^\mu(x) h^\rho(x) h^\rho(x) + \langle (\hat{\sigma}_{kh} - \langle \hat{\sigma}_{kh} \rangle_{\text{QM}})^2 \rangle_Q \tag{1.24 b}$$

Equations (1.21-24) essentially shows that in the language of ordinary QM the classical variables can be interpreted as approximate expressions of the corresponding quantum ones.

According to (1.24) the variance for the considered classical quantities results the sum of two terms. The first term, to which we shall refer as the *classical variance*, depends ony on weight functions and on the constants $\mu$ and $\nu_k$, that occur in the expression (1.20) and have the dimension of masses. The second one is simply the ordinary *quantum variance*. The constants $\mu$ and $\nu_k$ are nnecessary to make certain mathematical expressions well defined, but can be also taken 0 in the finul results- In the last case no new unknown would be introduced in the theory.

Obviously, if the variances as given by (1.24) are on the whole negligible, the classical quantities coincide with the expectation value of the corresponding quantum operators; then they evolve deterministically according some differential equations and we are in the field of Classical Physics. In such situation if in addition the quantum state is such that $\langle \hat{T}_f \rangle_Q$, $\langle \sigma_{1h} \rangle$, $\langle \sigma_{2h} \rangle$, ... can be decomposed in different specific contributions (as different phases, densities of various chemical components, fluxes of ions etc.) plus some interation terms a similar decomposition can be defined for the classical corresponding quantities.

If the classical variances are negligible, as we usually we assume, but the quantum ones are important, we are in the domain of application of ordinary Quantum Theory for observable for which an effective measurement procedure is provided.

If even the classic variance were relevant, we should be in presence of specific effects of the present modified theory. Presently, however, we do not know of any such case and In sec. 5 from consideretions in particular on the lifetime of the unstable particles and the fact that the standard model is consistent up to 0.1 %, we conclude that e.g. the upper bound $\mu < 0.1 \, eV$ has to hold.

Furtherly, if we integrate (1.20) with respect to some argument, we obtain the corresponding expression in the residual arguments

$$\int d\vartheta_{f_1} \ldots d\vartheta_{f_m} \, d\sigma_{1h_{11}} \ldots d\sigma_{1h_{1s_1}} \ldots \; p\left(\vartheta_{f_1}, \ldots \vartheta_{f_m}, \ldots \; ; \; \sigma_{1h_{11}}, \ldots \sigma_{1h_{1s_1}}, \ldots \; ; \; t_b, t_a ; \widehat{W}_a\right) =$$
$$= p\left(\vartheta_{f_{m+1}}, \vartheta_{f_{m+2}}, \ldots \; ; \; \sigma_{1h_1 r_1+1}, \sigma_{1h_1 r_1+2}, \ldots ; t_b, t_a ; \widehat{W}_a\right). \quad (1.25)$$

In particular the integration on all argument has as result 1.

Finally let us assume $t_a = -\infty$ and that $f_1, f_2, \cdots f_m$ and $h_{11}, h_{12}, \cdots h_{1s_1}, h_{21} \ldots$ have supports in the spacetime region $t < t_c < t_b$, while $f_{m+1}, f_{m+2}, \cdots$ and $h_{11}, h_{12}, \cdots h_{1s_1}, h_{21} \ldots$ have supports in the region $t_c < t < t_b$. Let us even suppose that $\vartheta_{f_1}, \cdots \vartheta_{f_m}$, $\sigma_{1h_{11}}, \ldots \sigma_{1h_{1s}}, \sigma_{21}, \ldots$ have actually achieved the values $\bar{\vartheta}_{f_1}, \cdots \bar{\vartheta}_{f_m} \bar{\sigma}_{1f_{11}}, \ldots \bar{\sigma}_{1h_{1s}}, \bar{\sigma}_{21}, \ldots$
We shall show that the conditional probability

$$p\left(\bar{\vartheta}_{f_1}, \ldots \bar{\vartheta}_{f_m} ; \; \bar{\sigma}_{1h_{11}}, \ldots \bar{\sigma}_{1h_{1s_1}} ; \bar{\sigma}_{21} \ldots \; \middle| \; \vartheta_{f_{m+1}}, \vartheta_{f_{m+2}}, \ldots ; \; \sigma_{1h_1 s_1+1}, \sigma_{1h_1 s_1+2}, \ldots ; \ldots\right) =$$

$$= \frac{p\left(\bar{\vartheta}_{f_1}, \ldots \bar{\vartheta}_{f_m}, \vartheta_{f_{m+1}}, \vartheta_{f_{m+2}}, \ldots \; ; \; \bar{\sigma}_{1h_{11}} \ldots \bar{\sigma}_{1h_{1s_1}}, \sigma_{1h_1}, \sigma_{1h_1 s_1+2}, \ldots ; \bar{\sigma}_{21}, \ldots ; t_b, -\infty ; \widehat{W}_{\text{in}}\right)}{p\left(\bar{\vartheta}_{f_1}, \ldots ; \bar{\sigma}_{h_{11}}, \ldots ; t_c, -\infty; \widehat{W}_{\text{in}}\right)} \quad (1.26)$$

Can be put in the form (1.20) with $a \to c$ and an appropriate definition of $\widehat{W}_c$. This is the form that the state reduction takes in the present perspective!

Finally let us come back to our cosmological model and in particular to the fluctuations in the CMB. For the theoretical aspect we shall refer to standard text books, like refs. [15-17] and try to conform as much as possible to the ref. [15] notations. However, note that we shall assume the metric and so the coefficient $a(t)$ adimentional. For the data we shall mainly refer to the Planck collaboration [18]. In the context of the ΛCDM model CMB fluctuations are normally explained as the result of the amplification of quantum fluctuations in the gravitational field during the inflation. Such fluctuations can be classified in three different types, scalar, vector and tensor modes. Vector fluctuations decay rapidly and are not usually taken into account.

The scalar fluctuations are directly driven by the inflaton field and so they depend on the choice of the potential $V(\varphi)$. When an appropriate $V(\varphi)$ is assumed their consideration enables to reproduce with good accuracy the angular correlations of the temperature-temperature futtuations (TT), the temperature-E type polarization fluctuations (TE) and the EE polarization fluctuations.

The tensorial fluctuations should be related to the true quantum character of $\hat{\gamma}_{\rho\sigma}$. They would imply polarization correlations of the type BB which are not affected by the scalar mode and should clearly be distinguished from them. However such type of polarization is not observed.

For our model and the potential (1.10) the correct scalar fluctuations are obtained for
$$\kappa = 8.38 \times 10^{12} \text{ GeV}, \qquad \lambda = 1.05 \times 10^{-15}. \qquad (1.27)$$
However, the ratio of the intensity of the tensorial to the scalar fluctuations is obtained for such values $r(q) \equiv \frac{4\,|D_q^0|^2}{|R_q^0|^2} = 0.086$, while a combination of the Planck most recent results with those of Bicep KEK leads to the bound $r < 0.032$ at 95% C. L [19].

Such difficulty is presently common to all the models based on inflation practically for any reasonable choice of the potential and could to suggest that Gravity is a purely classical phenomenon. In our framework such a perspective could be attained in the simplest way by making the somewhat *ad hoc* but consistent assumption $\hat{h}_{\rho\sigma} \equiv 0$ and so replaicing $\hat{\gamma}_{\rho\sigma}$ with $\bar{g}_{\rho\sigma}$ everywhere .

The layout of the paper is as follows. In sec. 2 we introduce the basic postulate for the probability (1.20) in the usual operator formalism and show that eqs. (1.24-26) are satisfied. In sec. 3 the postulate is restated in the language of the functional integral, it is shown that the definition (2.1) is positive and a more transparent expression for the probability is derived. Sec. 4 is devoted ti a discussion of the fluctuations on CMB on the line of ref. [15]. *In Sec.* 5 an upper bound limited to the constant $\mu$ is established. Finally in the appendix a numerical table concerning resolution of the system (4,10) satisfying the initial condition (4.2) and the determination of the time of exit from the horizon is reported.

## 2. Basic postulate

To introduce our basic postulate we shall start from a generating functional. First let us define two *superoperators* acting on the trace class operators of the Hilbert space of the universe,
$$\mathcal{K}(x\,;\,\xi(x),\,\eta_1(x),\ldots\eta_r(x)\,)\widehat{W} =$$
$$\frac{1}{2}\Big\{i\big(\xi^{\mu\nu}(x)\,\widehat{T}_{\mu\nu}(x) + \sum_{k=1}^{r}\eta_k^\mu(x)\widehat{J}_{k\mu}(x)\big) - \frac{\mu^4}{2}\,[-\hat{\gamma}(x)]^{-\frac{1}{2}}\,\hat{\gamma}_{\mu\rho}(x)\hat{\gamma}_{\nu\sigma}(x)\,\xi^{\mu\nu}(x)\,\xi^{\rho\sigma}(x) - $$
$$-\sum_{k=1}^{r}\frac{v_k^3}{2}\,[-\hat{\gamma}(x)]^{-\frac{1}{2}}\,\hat{\gamma}_{\mu\rho}(x)\hat{\gamma}_{\nu\sigma}(x)\,\eta_k^\mu(x)\eta_k^\rho(x)\,,\,\,\widehat{W}\Big\} \qquad (2.1)$$
and

$$\mathcal{G}[t_b, t_a; \xi, \eta_1, \ldots \eta_r] = T \exp \int_{t_a}^{t_b} d^4x \; \mathcal{K}(x; \xi(x), \eta_1(x), \ldots \eta_r(x)), \qquad (2.2)$$

where, as usual, $\{\hat{A}, \hat{B}\} = \hat{A}\hat{B} + \hat{B}\hat{A}$ and T denotes the time ordering prescription. We suppose again the quantities $\xi^{\mu\nu}(x)$ and $\eta_k^\mu(x)$ to transform as a tensor and a vector density and the complicate $\hat{\gamma}_{\mu\nu}$ arrangement in (2.1) is to make (2.2) coordinate independent.

Then we assume

$$p(\vartheta_{f_1}, \vartheta_{f_2}, \ldots; \sigma_{1h_{11}}, \sigma_{1h_{12}}, \ldots; \ldots; \sigma_{2h_{r1}}, \sigma_{2h_{r2}}, \ldots; t_b, t_a; \widehat{W}_a) =$$
$$\frac{1}{(2\pi)^{n+n_1+\ldots n_r}} \int dk_1 \cdots dk_n \, dp_{11} \ldots dp_{1n_1} \ldots dp_{r1} \ldots dp_{rn_r} \exp\left(i \sum_s k_s \vartheta_{f_s} + i \sum_{ks} p_{ks} \sigma_{h_{ks}}\right) \cdot$$
$$\cdot \operatorname{Tr}\{\mathcal{G}[t_b, t_a; \sum_j k_j \, f_j + \sum_{ks} p_{ks} h_{ks}] \widehat{W}_a\}. \qquad (2.3)$$

Such expression is inspired by the theory of continuous monitoring of refs. [8-9]. However a direct justification comes from its properties as described in the preceding section.

For simplicity from now on we shall keep trace only of the most problematic energy momentum tensor, the explicit inclusion of the currents being trivial.

As we mentioned by rewriting (2.3) in the language of the functional integral we shall see In the next section that, under the initial condition (1.11) and some additional assumptions on weight functions, the integral in the $k$'s are convergent and the resulting quantity positive. Then we may note that, due to the usual equation $\int_{-\infty}^{\infty} d\vartheta \exp(ik\vartheta) = 2\pi \delta(k)$, eq. (1.25) is trivially satisfied and, since

$$\mathcal{G}[t_b, t_a; 0] = 1, \qquad (2.4)$$

we have in paticular

$$\int d\vartheta_{f_1} d\vartheta_{f_2}, \cdots d\vartheta_{f_p} \, p(\vartheta_{f_1}, \vartheta_{f_2}, \cdots \vartheta_{f_n}) = 1. \qquad (2.5)$$

Furthermore, from the general equation

$$\langle \vartheta_{f_1} \vartheta_{f_2} \cdots \vartheta_{f_n} \rangle = \int d\vartheta_{f_1} d\vartheta_{f_2}, \cdots d\vartheta_{f_n} \, p(\vartheta_{f_1}, \vartheta_{f_2}, \cdots \vartheta_{f_n}) \, \vartheta_{f_1} \vartheta_{f_2} \cdots \vartheta_{f_n} =$$
$$= i\frac{\partial}{\partial k_1} i\frac{\partial}{\partial k_2} \cdots i\frac{\partial}{\partial k_n} \operatorname{Tr}\{\mathcal{G}[t_b, -\infty; \sum_j k_j \, f_j(t)] \widehat{W}_{\text{in}}\}_{k_1 = k_2 = \cdots k_n = 0} \qquad (2.6)$$

(1.21, 24) follow immediately. Finally from the obvious property

$$\mathcal{G}[t_b, -\infty; \xi] = \mathcal{G}[t_b, t_c; \xi] \mathcal{G}[t_c, -\infty; \xi], \qquad (2.7)$$

under the assumptions made we have (1.26)

$$p(\bar{\vartheta}_{f_1}, \bar{\vartheta}_{f_2}, \cdots \bar{\vartheta}_{f_m} | \vartheta_{f_{m+1}}, \vartheta_{f_{m+2}}, \cdots \vartheta_{f_n}) =$$
$$= \frac{1}{(2\pi)^{n-m-1}} \int dk_{m+1} dk_{m+2} \cdots dk_n \, \exp\left(i \sum_{j=m+1}^n k_j \vartheta_{f_j}\right) \operatorname{Tr}\{\mathcal{G}[t_b, t_c; \sum_{j=m+1}^n k_j f_j] \widehat{W}_c\},$$
$$\qquad (2.8)$$

whith

$$\widehat{W}_c = \frac{\frac{1}{(2\pi)^m} \int dk_1 dk_2 \cdots dk_m \, \exp\left(i \sum_{j=1}^m k_j \bar{\vartheta}_{f_j}\right) \{\mathcal{G}[t_c, -\infty; \sum_{j=1}^m k_j f_j] \widehat{W}_{\text{in}}\}}{\frac{1}{(2\pi)^n} \int dk_1 dk_2 \cdots dk_n \, \exp\left(i \sum_{j=1}^m k_j \bar{\vartheta}_{f_j}\right) \operatorname{Tr}\{\mathcal{G}[t_c, -\infty; \sum_{j=1}^m k_j f_j] \widehat{W}_{\text{in}}\}}, \qquad (2.9))$$

which is trace 1 and positive.

## 3. Functional Integral

To go on, it is convenient to rewrite (2.2) In the language of the functional integral.
We have

$\langle h_{\mu\nu b}, \chi_b, \phi_{1b}, \cdots ; t_b | \mathcal{G}[t_b \; t_a \; ; \xi] \widehat{W}_a | h'_{\mu\nu b}, \chi'_b, \phi'_{1b}, \ldots, \cdots ; t_b \rangle =$

$\int \mathcal{D}_{t_a} h_{\mu\nu a} \mathcal{D}_{t_a} \chi_a \, \mathcal{D}_{t_a} \phi_{1a} \cdots \int \mathcal{D}_{t_a} h'_{\mu\nu a} \mathcal{D}_{t_a} \chi'_a \, \mathcal{D}_{t_a} \phi'_{1a} \cdots \, \langle h_{\mu\nu a}, \chi_a, \phi_{1a}, \ldots | \widehat{W}_a | h'_{\mu\nu a} \; \chi'_a, \phi'_{1a}, \cdots \rangle$

$\int_{h_{\mu\nu a},\chi_a,\phi_{1a},\cdots}^{h_{\mu\nu b},\chi_b,\phi_{1b},\cdots} \mathcal{D} h_{\mu\nu} \mathcal{D} \chi \, \mathcal{D} \phi_1 \cdots \int_{h'_{\mu\nu a},\chi'_a,\phi'_{1a},\cdots}^{h'_{\mu\nu b},\chi'_b,\phi'_{1b},\cdots} \mathcal{D} h'_{\mu\nu}{}' \mathcal{D} \chi' \, \mathcal{D} \phi'_1 \cdots$

$\exp \left\{ \int_{t_a}^{t_b} d^4 x \left[ -\frac{i}{2} \xi^{\mu\nu}(x) \, [T_{\mu\nu}(\gamma_{\rho\sigma}, \varphi, \phi_1(x), \cdots) + T_{\mu\nu}(\gamma'_{\rho\sigma}, \varphi', \phi'_1(x), \cdots)] \right. \right.$

$\left. - \frac{\mu^4}{8} \left[ (-\gamma(x))^{-\frac{1}{2}} \gamma_{\mu\rho}(x) \, \gamma_{\nu\sigma}(x) + (-\gamma'(x))^{-\frac{1}{2}} \gamma'_{\mu\rho}(x) \, \gamma'_{\nu\sigma}(x) \right] \xi^{\mu\nu}(x) \, \xi^{\rho\sigma}(x) \right] +$

$\left. + i \left( S_T[\gamma_{\mu\nu}, \varphi, \phi_1, \ldots] - S_T[\gamma'_{\mu\nu}, \varphi', \phi'_1, \cdots] \right) \right\},$  (3.1)

where $S_T$ stays for the total action as defined in eqs. (1.7-9).

The above expression is obviously thought defined as the limit from a finite lattice, $\mathcal{D}\gamma_{\mu\nu}, \mathcal{D}\varphi, \mathcal{D}\phi_s$ and $\mathcal{D}_t\gamma_{\mu\nu}, \mathcal{D}_t\varphi, \mathcal{D}_t\phi_s$ denote the Feynman measures, given by

$$\mathcal{D}\phi_s \propto \prod_{j=1}^N d\phi_s(x_j) \qquad (3.2)$$

$$\mathcal{D}_t\phi_s \propto \prod_{j=1}^M d\phi_s(t, x_j) \qquad (3.3)$$

up to factors which depend on the field types. For the gauge and in particular the gravitational fields, a gauge is supposed to be chosen and the corresponding Faddeev Popov determinant [20] included. The lower and upper indices in the integrals on the third line of (3.1) want indicate that the values of the fields at the two border $t = t_a$ and $t = t_b$ are kept fixed on the specified values. Finally $|h_{\mu\nu}, \varphi, \phi_1, \ldots ; t \rangle$ denotes the simultaneous eigenvector corresponding to the specified eigenvalues of all Bose fields or bilinear expressions of the Fermi fields in all points of the lattice at the time $t$

$$\hat{\phi}_s(t, x_j) | h_{\mu\nu}, \varphi, \phi_1, \ldots ; t \rangle = \phi_s(t, x_j) | h_{\mu\nu}, \varphi, \phi_1, \ldots ; t \rangle \qquad (3.4)$$

etc. They are assumed to be normalized according the equation

$$\langle h_{\mu\nu}, \varphi, \phi_1, \cdots ; t | h_{\mu\nu}, \varphi', \phi'_1, \cdots ; t \rangle = \delta[h_{\mu\nu} - h'_{\mu\nu}] \delta[\varphi - \varphi'] \delta[\phi_1 - \phi'_1] \cdots , \qquad (3.5)$$

where $\delta[\phi_s - \phi'_s]$ etc. stay for the functional space $\delta$- functions defined by

$$\int \mathcal{D}_t \phi_s \, \delta[\phi_s - \phi'_s] \, F[\phi_s] = F[\phi'_s], \quad \text{etc.} \qquad (3.6)$$

More rigorously but less intuitively one could use the formalism of the olomorphic representation (see [20]).

Now, let us set $t_a = -\infty$ and $\widehat{W}_{in} = |0\rangle\langle 0|$, as we have assumed. Then for a known property of the Feynman integral, expression (3.1) can be written up to a renormalization constant simply setting all $h_{\mu\nu a} = h'_{\mu\nu a}$, $\chi_a = \chi'_a$, $\phi_{1a} = \phi'_{1a}$, ... equal to 0. Omitting explicit reference to the currents, (2.3) becomes

$p(\vartheta_{f_1}, \vartheta_{f_2}, \cdots \vartheta_{f_n}; t_b, -\infty) = \frac{1}{(2\pi)^n} \int dk_1 dk_2 \cdots dk_n \, \exp\left(i \sum_j k_j \vartheta_{f_j}\right)$

$C \int \mathcal{D}_{t_b} h_{\mu\nu b} \mathcal{D}_{t_b} \chi_b \, \mathcal{D}_{t_b} \phi_{1b} \cdots \int_{0,0,0,\cdots}^{h_{\mu\nu b},\chi_b,\phi_{1b},\cdots} \mathcal{D} h_{\mu\nu} \mathcal{D} \chi \, \mathcal{D} \phi_1 \cdots \int_{0,0,0,\cdots}^{h_{\mu\nu b},\chi_b,\phi_{1b},\cdots} \mathcal{D} h'_{\mu\nu} \mathcal{D} \chi' \, \mathcal{D} \phi'_1 \cdots$

$$\exp\left\{\int_{-\infty}^{t_b} d^4x \left[-\frac{i}{2}\sum_j k_j f_j^{\mu\nu}(x) \left[T_{\mu\nu}(\gamma_{\rho\sigma}, \varphi, \phi_1(x), \cdots) + T_{\mu\nu}(\gamma'_{\rho\sigma}, \varphi', \phi'_1, \cdots)\right] - \right.\right.$$
$$\left. - \frac{\mu^4}{8}\left[[-\gamma(x)]^{-1/2}\gamma_{\mu\rho}(x)\gamma_{\nu\sigma}(x) + [-\gamma'(x)]^{-1/2}\gamma'_{\mu\rho}(x)\gamma'_{\nu\sigma}(x)\right] k_i k_j f_i^{\mu\nu}(x) f_j^{\rho\sigma}(x)\right] +$$
$$\left. + i\left(S_T[\gamma_{\mu\nu}, \varphi, \phi_1, \ldots] - S_T[\gamma'_{\mu\nu}, \varphi', \phi'_1, \cdots]\right)\right\} =$$
$$= C \int \mathcal{D}_{t_b} h_{\mu\nu b} \mathcal{D}_{t_a} \chi_b \, \mathcal{D}_{t_b} \phi_{1b} \cdots \int_{0,0,0,\ldots}^{h_{\mu\nu b}, \chi_b, \phi_{1b}, \cdots} \mathcal{D}h_{\mu\nu} \mathcal{D}\chi \, \mathcal{D}\phi_1 \cdots \int_{0,0,0,\ldots}^{h_{\mu\nu b}, \chi_b, \phi_{1b}, \cdots} \mathcal{D}h'_{\mu\nu} \mathcal{D}\chi' \, \mathcal{D}\phi'_1 \cdots$$

$$\frac{1}{\mu^2 \sqrt{(2\pi)^p \det\left[\frac{M(\gamma_{\mu\nu}) + M(\gamma'_{\mu\nu})}{2}\right]_{ij}}} \exp\left\{-\frac{1}{2\mu^4}\sum_{ij}\left[\frac{M(\gamma_{\mu\nu}) + M(\gamma'_{\mu\nu})}{2}\right]_{ij}^{-1} \left[\vartheta_{f_i} - \right.\right.$$

$$\left. - \frac{T_{f_i}(\gamma_{\mu\nu}, \varphi, \phi_1(x), \cdots) + T_{f_i}(\gamma'_{\mu\nu}, \varphi', \phi'_1(x), \cdots)}{2}\right]\left[\vartheta_{f_j} - \frac{T_{f_j}(\gamma_{\mu\nu}, \varphi, \phi_1(x), \cdots) + T_{f_j}(\gamma'_{\mu\nu}, \varphi', \phi'_1(x), \cdots)}{2}\right] +$$

$$\left. + i\left(S_T[\gamma_{\mu\nu}, \varphi, \phi_1, \ldots] - S_T[\gamma'_{\mu\nu}, \varphi', \phi'_1, \cdots]\right)\right\}, \quad (3.7)$$

where
$$C^{-1} =$$
$$\int \mathcal{D}_{t_b} h_{\mu\nu b} \mathcal{D}_{t_a} \chi_b \, \mathcal{D}_{t_b} \phi_{1b} \cdots \int_{0,0,0,\ldots}^{h_{\mu\nu b}, \chi_b, \phi_{1b}, \cdots} \mathcal{D}h_{\mu\nu} \mathcal{D}\chi \, \mathcal{D}\phi_1 \cdots \int_{0,0,0,\ldots}^{h_{\mu\nu b}, \chi_b, \phi_{1b}, \cdots} \mathcal{D}h'_{\mu\nu} \mathcal{D}\chi' \, \mathcal{D}\phi'_1 \cdots$$
$$\exp\left\{i\left(S_T[\gamma_{\mu\nu} \varphi, \phi_1, \ldots] - S_T[\gamma'_{\mu\nu} \varphi', \phi'_1, \cdots]\right)\right\} \quad (3.8)$$
and
$$M_{ij}(\gamma_{\mu\nu}) = \frac{\mu^4}{2} \int d^4x \, [-\gamma(x)]^{-1/2} \gamma_{\mu\rho}(x) \gamma_{\nu\sigma}(x) f_i^{\mu\nu}(x) f_j^{\rho\sigma}(x). \quad (3.9)$$

Obviously in order the integral (3.7) to be convergent the weight function have to be chosen in such a way that the matrix $M_{ij}$ is positive. In particular the classic variance defined in (1.24) must be positive. The zero order approximation of (3.9) is

$$M_{ij}^{(0)} = \frac{\mu^4}{2} \int d^4x \, a^{-3}(t) \, \bar{g}_{\mu\rho}(x) \, \bar{g}_{\nu\sigma}(x) f_i^{\mu\nu}(x) f_j^{\rho\sigma}(x). \quad (3.10)$$

Corrections can be obtained by expanding (3.9) in $h_{\mu\nu}(x)$, but they have the same positivity and can be assumed small in the following. Then note that eq. (3.7) cannot give any information on the values of the component $t_{0j}$ of the classical energy moment tensor or similarly of $J_{k0}$ of the classical current $k$. However such quantities are provided in terms of the other components by the conservation equations (1.5).

Now let us come back to the general properties of the quantity defined by (2,3) or equivalently by (3.7). We have still to prove that it is positive, what is essential in order it can be interpreted as a probability. To this aim let use the Euler relation

$$\exp\left\{i\left(S_T[\gamma_{\mu\nu} \varphi, \phi_1, \ldots] - S_T[\gamma'_{\mu\nu} \varphi', \phi'_1, \cdots]\right)\right\} =$$
$$= \cos\left\{S_T[\gamma_{\mu\nu} \varphi, \phi_1, \ldots] - S_T[\gamma'_{\mu\nu} \varphi', \phi'_1, \cdots]\right\} +$$
$$+ i \sin\left\{S_T[\gamma_{\mu\nu} \varphi, \phi_1, \ldots] - S_T[\gamma'_{\mu\nu} \varphi', \phi'_1, \cdots]\right\} \quad (3.11)$$

and note that the second term does not give contribution in (3.7). In fact this is antisymmetric in the exchange of all the primed and the unprimed arguments, while the remaining part of the integrand is symmetric. On the other side the actions are dominated by the classical trajectories of the fundamental fields which are identical in this case since the primed and unprimed variables are equal at their extremes.. Then the first term is dominated by the vanishing value of its argument and so it is positive.

### 4. Quantum fluctuations. Fluctuations in the Cosmic Microwave Background.

Let us now consider the problem of he fluctuations of the CMB, that it is one of the main constraints on cosmological models, in our context. We find that the standard methods fpr their evaluation can be applied independently of the choice of the origin of time.

Let us go back to eqs. (1.7-10) and consider an analogous splitting of the quantum energy-momentum tensor in the pure inflaton contribution and the residual part

$$T_{\mu\nu}(\gamma_{\rho\sigma}; \varphi, \phi_1, \dots) = T_{\varphi\,\mu\nu}(\gamma_{\rho\sigma}; \varphi) + T_{\phi\,\mu\nu}(\gamma_{\rho\sigma}; \varphi, \phi_1, \dots) \,, \qquad (4.1)$$

with

$$T_{\varphi\,\mu\nu}(\gamma_{\rho\sigma}; \varphi) = -\partial_\mu\varphi\,\partial_\nu\varphi - \gamma_{\mu\nu}(\tfrac{1}{2}\gamma^{\rho\sigma}\partial_\rho\varphi\,\partial_\sigma\varphi + V(\varphi)) \,. \qquad (4.3)$$

We have

$$T_{\varphi\,\mu\nu}(\gamma_{\rho\sigma}; 0) = -\gamma_{\mu\nu}\,\frac{\kappa^4}{4\lambda} \,, \qquad (4.4)$$

while the second term in (4.1) vanishes, when all fields except $\varphi$ and $\gamma_{\mu\nu}$, are set equal to 0, a part minor contributions from other scalar fields associated to symmetry breakings, like the Higgs field of the standard model. Notice that according to (1.27) $\frac{\kappa^4}{4\lambda}\sim 10^{66}$ GeV$^4$ while the corresponding quantity for the Higgs is of the order $10^9$ GeV$^4$.

Due to the assumptions (1,11-14), sufficiently far in the past, the second term in (4.1) shall be negligible with respect to the first one. A similar circumstance occurs for the action $S_\phi[\gamma_{\mu\nu}, \varphi, \phi_1, \phi_2, \cdots]$ in equation (1.7). Then, if the support of all the weight functions are far in the past, we can eliminate a factor between the numerator and the normalization constant in (3.7) and rewrite the latter as

$$p(\vartheta_{f_1}, \vartheta_{f_2}, \cdots \vartheta_{f_n}; t_b, -\infty)$$
$$= C' \int \mathcal{D}_{t_b} h_{\mu\nu b} \mathcal{D}_{t_a} \chi_{b\,t_b} \int_{0,0}^{h_{\mu\nu b}, \chi_b} \mathcal{D}h_{\mu\nu}\mathcal{D}\chi \int_{0,0}^{h_{\mu\nu b}, \chi_b} \mathcal{D}h'_{\mu\nu}\mathcal{D}\chi'$$

$$\frac{1}{\mu^2\sqrt{(2\pi)^p \det\left[\frac{M(\gamma_{\mu\nu})+M(\gamma'_{\mu\nu})}{2}\right]_{ij}}} \exp\left\{\left[-\frac{1}{2\mu^4}\Sigma_{ij}\frac{M_{ij}(\gamma_{\mu\nu})+M_{ij}(\gamma'_{\mu\nu})}{2}\left(\vartheta_{f_i}-\right.\right.\right.$$

$$\left.\frac{T_{\varphi f_i}(\gamma_{\rho\sigma}; \varphi) + T_{\varphi f_i}(\gamma'_{\rho\sigma}; \varphi')}{2}\right)\left(\vartheta_{f_j}--\frac{T_{\varphi f_i}(\gamma_{\rho\sigma}; \varphi)+T_{\varphi f_j}(\gamma'_{\rho\sigma}; \varphi')}{2}\right)\right] + i(S_\gamma[\gamma_{\mu\nu}] + S_\varphi[\gamma_{\mu\nu}, \varphi]) -$$

$$- i(S_\gamma[\gamma'_{\mu\nu}] + S_\varphi[\gamma'_{\mu\nu}, \varphi'])\Bigg\} \,, \qquad (4.5)$$

with

$$(C')^{-1} = \int \mathcal{D}_{t_b} h_{\mu\nu b}\mathcal{D}_{t_a}\chi_{b\,t_b} \int_{0,0}^{h_{\mu\nu b}, \chi_b} \mathcal{D}h_{\mu\nu}\mathcal{D}\chi \int_{0,0}^{h_{\mu\nu b}, \chi_b} \mathcal{D}h'_{\mu\nu}\mathcal{D}\chi'$$
$$\exp\{i(S_\gamma[\gamma_{\mu\nu}] + S_\varphi[\gamma_{\mu\nu}, \varphi]) - i(S_\gamma[\gamma'_{\mu\nu}] + S_\varphi[\gamma'_{\mu\nu}, \varphi'])\}, \qquad (4.6)$$

in which only the inflaton field and the quantum metric appear.

Now let us consider the Euler equations for the reduced action

$$S_\gamma[\gamma_{\mu\nu}] + S_\varphi[\gamma_{\mu\nu}, \varphi] \ . \tag{4.7}$$

These are

$$\partial_\rho(\sqrt{-\gamma}\,\gamma^{\rho\sigma}\partial_\sigma\varphi) - \sqrt{-\gamma}\,V'(\varphi) = 0 \tag{4,8}$$

(Where the apex stays for derivative with respect to $\varphi$) and

$$R_{\mu\nu}(\gamma_{\rho\sigma}) - \tfrac{1}{2}\gamma_{\mu\nu} R(\gamma_{\rho\sigma}) = -8\pi G\, T_{\varphi\,\mu\nu}(\gamma_{\rho\sigma}; \varphi)). \tag{4.9}$$

Such equations can be interpreted as specifying the dominant field configurations in the functional integral or, equivalently, as Heisenberg picture field operator equations.

At zero order in $\chi$ and $h_{\mu\nu}$, setting $\varphi(x) = \bar\varphi(t)$ and $\gamma_{\rho\sigma}(x) = \bar g_{\rho\sigma}(t)$ we obtain equations (1.15) and (1.16), which we write again

$$\ddot\varphi(t) + 3\tfrac{\dot a}{a}\,\dot\varphi(t) + V'(\varphi(t)) = 0 \ , \tag{4.10}$$

$$\frac{\dot a(t)}{a(t)} = \sqrt{\frac{8\pi G}{3}\left(\tfrac{1}{2}\dot\varphi^2(t) + V(\varphi(t))\right)} \ . \tag{4.11}$$

The asymptotic solutions of such system which vanish for $t \to -\infty$ are

$$\varphi = c\,e^{\alpha t}, \qquad a = \bar a\,e^{\bar H t} \ , \tag{4.12}$$

where

$$\bar H = \sqrt{\frac{8\pi G}{3}\frac{\kappa^4}{4\lambda}} \ , \qquad \alpha = \tfrac{1}{2}\left(-3\bar H + \sqrt{9\bar H^2 + 4\kappa^2}\right) \ , \tag{4.13}$$

$\bar H$ being the $t \to -\infty$ value of the *Huble-Lemaitre coefficient* $H(t) = \tfrac{\dot a}{a}$ and $c$ amd $\bar a$ two non significant arbitrary coeficients. In fact $\bar a$ can be reabsorbed in a redefinition of the space coordinates, while we can always put $c = c'e^{-\alpha t_1}$ and so the value of $c$ fixes only the origin of the time axis. We shall take $c = v$, with $v = \tfrac{\kappa}{\sqrt\lambda}$ (the second zero of $V'(\varphi)$ for which $V(\varphi)$ is minimum and vanishes). As we shall see, with such choice inflation ends at a time about $t = 0$ while a typical time for one e-folding is $\tfrac{1}{\bar H} \sim 5 \times 10^{-15}$ GeV$^{-1}$ $\sim 10^{-39}$ sec .

Quantitatively, for finite $t$ the system (4.10-11) can be solved numerically (see Fig. 1-3 and the table reported in Appendix). Qualitatively $\varphi(t)$ and $a(t)$ start from the form (4.12) and continue to increase until $\varphi(t)$ crosses the value $v$ and then it decays to such value with damped oscillations while $a(t)$ goes to a constant (Fig. 1, 2). The time $t_I$ at which this happen can be considered as the end of the inflation and in some way the origin of our world. With the values (1.27) we can take $t_I \sim 5 \times 10^{-12}$ GeV$^{-1}$.

To study small fluctuation around the above zero order solutions, first let us take the gauge of Newton and write (see ref. [15], Cap. 5).

$$h_{00} = -2\Phi, \qquad h_{0j} = a\,G_j, \qquad h_{ij} = a^2\,(-2\,\Psi\delta_{ij} + D_{ij}) \ , \tag{4.14}$$

with

$$\partial_j G_j = 0, \qquad \partial_i D_{ij} = 0, \qquad D_{ij} = D_{ji}, \qquad D_{jj} = 0 \ . \tag{4.15}$$

Then at the first order in $\chi$ e $h_{\rho\sigma}$ eq. (4.8) becomes

$$\ddot{\hat\chi} + 3\tfrac{\dot a}{a}\dot{\hat\chi} + V''(\bar\varphi)\hat\chi - \tfrac{\nabla^2}{a^2}\hat\chi = -2\,\hat\Phi\,V'(\bar\varphi) + \left(\dot{\hat\Phi} + 3\dot{\hat\Psi}\right)\dot{\bar\varphi}\,, \tag{4.16}$$

where obviously the hat indicates that now the corresponding quantities have to be interpreted as perators.

In the same approximation eq. (4.9) can be reorganized in three set of independent equations (see ref. [15, chaps. 5 e 10]):

1) scalar mode
$$\partial_i \partial_j (\widehat{\Phi} - \widehat{\Psi}) = 0 \tag{4.17}$$
$$\dot{a}\, \partial_i \widehat{\Phi} + a\, \partial_i \dot{\widehat{\Psi}} = 4\pi G\, a\, \dot{\bar{\varphi}}\, \partial_i \hat{\chi} \tag{4.18}$$
$$\left(\dot{H} - \frac{\nabla^2}{a^2}\right) \widehat{\Psi} = 4\pi G\, (\ddot{\bar{\varphi}}\hat{\chi} - \dot{\bar{\varphi}}\dot{\hat{\chi}}) \tag{4.19}$$

2) vector mode
$$\partial_i \left( a\, \dot{\widehat{G}}_j + 2\dot{a}\, \widehat{G}_j \right) = 0 \tag{4.20}$$

3) tensor mode
$$\ddot{\widehat{D}}_{ij} + 3\frac{\dot{a}}{a}\dot{\widehat{D}}_{ij} - \frac{\nabla^2}{a^2}\widehat{D}_{ij} = 0\,, \tag{4.21}$$

where $H(t) = \frac{\dot{a}}{a}$ as before.

Then eq. (4.20) has solutions of the form
$$\widehat{G}_j = \frac{c_j}{a^2}\,. \tag{4.22}$$

so $\widehat{G}_j$ decays fast as $t$ increases and it is neglected as a rule (see ref. [15], Chap. 5).

From eq. (4.17) it follows $\widehat{\Phi} = \widehat{\Psi}$. So eq. (4.16) becomes
$$\ddot{\hat{\chi}} + 3\frac{\dot{a}}{a}\dot{\hat{\chi}} + V''(\bar{\varphi})\hat{\chi} - \frac{\nabla^2}{a^2}\hat{\chi} = -2\widehat{\Psi} V'(\bar{\varphi}) + 4\dot{\widehat{\Psi}}\dot{\bar{\varphi}}\,. \tag{4.23}$$

and the general solution of the system of eqs. (4.23) and (4.18,19) can be written as
$$\hat{\chi}(x) = \int d^3 q\, \left(\hat{\alpha}(\boldsymbol{q})\chi_q(t) e^{i\boldsymbol{q}\cdot\boldsymbol{x}} + \hat{\alpha}^+(\boldsymbol{q})\chi_{\boldsymbol{q}}^*(t) e^{-i\boldsymbol{q}\cdot\boldsymbol{x}}\right), \tag{4.24}$$
$$\widehat{\Psi}(x) = \int d^3 q\, \left(\hat{\alpha}(\boldsymbol{q})\Psi_q(t) e^{i\boldsymbol{q}\cdot\boldsymbol{x}} + \hat{\alpha}^+(\boldsymbol{q})\Psi_{\boldsymbol{q}}^*(t) e^{-i\boldsymbol{q}\cdot\boldsymbol{x}}\right), \tag{4.25}$$

where $\chi_q(t)$ and $\Psi_q(t)$ respectively are defined by the equationsDigitare l'equazione qui.
$$\ddot{\chi}_q + 3\frac{\dot{a}}{a}\dot{\chi}_q + V''(\bar{\varphi})\chi_q + \frac{q^2}{a^2}\chi_q = -2\Psi_q V'(\bar{\varphi}) + 4\dot{\Psi}_q\dot{\bar{\varphi}} \tag{4.26}$$
$$\dot{\Psi}_q + \frac{\dot{a}}{a}\Psi_q = 4\pi G\, a\, \dot{\bar{\varphi}}\chi_q \tag{4.27}$$
$$\left(\dot{H} + \frac{q^2}{a^2}\right)\Psi_q = 4\pi G\, (\ddot{\bar{\varphi}}\chi_q - \dot{\bar{\varphi}}\dot{\chi}_q) \tag{4.28}$$

and the "constraints" $\hat{\alpha}(\boldsymbol{q})$ e la $(\hat{\alpha}^+(\boldsymbol{q}))$ have to be interpreted as operators.

Similarly the general solution of (4.21) is
$$D_{ij}(x) = \sum_{s=1}^{2} \int d^3 q\, \left(\widehat{\beta}_s(\boldsymbol{q})D_{ijqs}(t) e^{i\boldsymbol{q}\cdot\boldsymbol{x}} + \widehat{\beta}_s^+(\boldsymbol{q})D_{ijqs}^*(t) e^{-i\boldsymbol{q}\cdot\boldsymbol{x}}\right), \tag{4.29}$$

with $\widehat{\beta}_s(\boldsymbol{q})$ e $\widehat{\beta}_s^+(\boldsymbol{q})$ interpreted again as operators and having set
$$D_{ij\,qs}(t) = e_{sij}(\boldsymbol{q})\, D_q(t)\,, \tag{4.30}$$

with $e_{1ij}(q)$ and $e_{2ij}(q)$ polarization tensors satisfying the condition
$$e_{ij} = e_{ji}\,,\qquad e_{ii} = 0\,,\qquad q_i e_{ij} = 0\,,\qquad e_{sij} e_{s'}^{ij} = \delta_{ss'}$$

and $D_{qs}(t)$ solution of the equation
$$\ddot{D}_q + 3H\dot{D}_q + \frac{q^2}{a^2} D_q = 0\,. \tag{4.31}$$

Equations (4.26-28) coincides with (10.1.15-17) of ref. [15] and (4.31) with (10.2.3). Assuming the classic fluctuations (as defined in (1.24)) negligible (sect. 5) in comparison to the quantum fluctuations, they have the same consequences too.

Note that the quantity $\frac{q}{a(t)}$ is the modulus of the physical propagation vector. Such vector goes to infinity as $t \to -\infty$, while $H(t)$ goes to the constant $\bar{H}$. On the contrary, as $t$ increases, $\frac{q}{a(t)}$ decreases faster than $H(t)$. As a consequence a time $t_{1q}$ must exist such that $\frac{q}{a(t_{1q})} = H(t_{1q})$ and $\frac{q}{a(t)} > H(t)$ for $t < t_{1q}$. One usually says that for $t < t_{1q}$ the $q$ component is inside the horizon (the wave length $\frac{a}{q}$ is smaller than the characteristic expansion time of the Universe $\frac{1}{H(t)}$). On the contrary, for $t > t_{1q}$ we have $\frac{q}{a(t)} < H(t)$ and the component is out of the horizon.

For $t$ sufficiently smaller than $t_{1q}$, and so $\frac{q}{a} \gg H$, we can solve the system (4.26-28) up to order $\frac{q}{a}$. With an appropriate choice of the normalization, two independent solutions can be written as

$$\chi_q = \frac{1}{(2\pi)^{\frac{3}{2}} a(t) \sqrt{2q}} \exp\left(iq \int_t^{t_b} dt' \frac{1}{a(t')}\right) \qquad (4.32)$$

$$\Psi_q = \frac{4\pi G \, \dot{\bar{\varphi}}(t)}{(2\pi)^{\frac{3}{2}} q \sqrt{2q}} \exp\left(iq \int_t^{t_b} dt' \frac{1}{a(t')}\right) \qquad (4.33)$$

and their conjugate expressions. From the canonical commutation rules between $\hat{\chi}(x)$ e $\dot{\hat{\chi}}(x)$ then we have

$$[\hat{a}(\mathbf{q}), \hat{a}(\mathbf{q}')] = [\hat{a}^+(\mathbf{q}), \hat{a}^+(\mathbf{q}')] = 0, \qquad [\hat{a}(\mathbf{q}), \hat{a}^+(\mathbf{q}')] = \delta^3(\mathbf{q} - \mathbf{q}') \qquad (4.34)$$

and so $\hat{a}(\mathbf{q})$ and $\hat{a}^+(\mathbf{q})$ can be interpreted as destruction and creation operators of *inflatons*, the quanta of field $\chi(x)$..

In a similar way, under the corresponding assumptions and in the same approximation, two solutions of (4.30) are

$$D_q(t) = \frac{\sqrt{16\pi G}}{(2\pi)^{3/2} \sqrt{2q}\, a(t)} \exp\left(iq \int_t^{t_b} dt' \frac{1}{a(t')}\right) \qquad (4.35))$$

and the conjugate one. Again

$$[\hat{\beta}_s(\mathbf{q}), \hat{\beta}_{s'}(\mathbf{q}')] = [\hat{\beta}_s^+(\mathbf{q}), \hat{\beta}_{s'}^+(\mathbf{q}')] = 0, \qquad [\hat{\beta}_s(\mathbf{q}), \hat{\beta}_{s'}^+(\mathbf{q}')] = \delta_{ss'} \delta^3(\mathbf{q} - \mathbf{q}') \qquad (4.36)$$

and $\hat{\beta}_s(\mathbf{q})$ and $\hat{\beta}_s^+(\mathbf{q})$ result as creation and destruction operators of *gravitons* in the polarization state $s$.

Rules (4.34) and (4.36) imply small primordial fluctuations in the energy momentum tensor and in the e-m. current. Amplified as a consequence of the space expansion they can justify the existence of fluctuations in temperature and E type of polarization in the CMB (E and then B denoting appropriate combinations of *Stokes parameters* [15], sec. .7.4).

After the end of the inflation $t_I$ the scenario should be the following. The energy freed by the field $\bar{\varphi}(t)$ as it changes from the initial value 0 to $v$ turns out in creation of inflatons [17, chap. 2]. Inflatons decay in particle-antiparticle pairs and neutral particles. Then pairs annihilate in photons and other force quanta and primordial light nuclei are produced as the temperature progressively decreases. Lastly, when even electron-positron pairs have been annihilated, it is assumed that there remain a mixture of photons, neutrinos, dark matter and a small quantity of a *barion plasma* (made by electrons, protons and light nuclei) before vacuum energy or *quintessence* becomes important .

So to the period dominated by the inflaton field, a period follows dominated by radiation and matter. In the latter $a(t)$ increases as $t^{1/2}$ or $t^{2/3}$ while $H(t) = \frac{\dot{a}(t)}{a(t)}$ decreases as $1/t$. Then it must exist a second time $t_{2q}$ at which we have again $\frac{q}{a(t_{2q})} = H(t_{2q})$ and we are back inside the horizon.

The next evolution of the scalar fluctuations is determined by the quantity
$$R_q(t) = - \Psi_q(t) - H(t)\, \delta u_q(t) , \qquad (4.37)$$
where $\delta u_q(t)$ is the Fourier component of the velocity potential which results $\delta u_q(t) = -\frac{\chi_q(t)}{\dot{\bar\varphi}(t)}$ at the first order in our case. For $t_{1q} < t < t_{2q}$, $R_q(t)$ becomes independent of $t$, $R_q(t) \to R_q^0$, and it occurs in the initial conditions for the equations that determine the successive evolution of the Universe. In a similar way tensor fluctuations are determined by the limit value $D_q^0$ ot the quantity $D_q(t)$ that become constant in the same time interval.

In the so called "slow roll" approximation one finds
$$\lfloor R_q^0 \rfloor^2 = \frac{G\, H^2(t_{1q})}{4\pi^2 \epsilon(t_{1q})}\, q^{-3} \cong \lfloor N_S \rfloor^2\, q_R^{-3} \left(\frac{q}{q_R}\right)^{-4+n_S} , \qquad (4.38\ a)$$
$$\lfloor D_q^0 \rfloor^2 = \frac{G\, H^2(t_{1q})}{\pi^2}\, q^{-3} \cong \lfloor N_T \rfloor^2\, q_R^{-3} \left(\frac{q}{q_R}\right)^{-3+n_T} , \qquad (4.38\ b)$$
where $q_R$ (denoted as $k_R$ in [15]) is a largely arbitrary constant with the dimension of an energy usually taken as [15], Chap. 7.10.
$$q_R = 0.05\ \text{Mpc}^{-1} = 3.193 \times 10^{-40}\ \text{GeV} , \qquad (4.39)$$

$$\lfloor N_S \rfloor^2 = \frac{G\, H^2(t_{1q_R})}{4\pi^2 \epsilon(t_{1q_R})} , \qquad \lfloor N_T \rfloor^2 = \frac{G\, H^2(t_{1q_R})}{\pi^2} \qquad (4.40)$$
and
$$n_S = 1 - 4\epsilon(t_{1q_R}) - 2\delta ,(t_{1q_R}) \qquad n_T = -2\varepsilon(t_{1q_R}) , \qquad (4.41)$$
with
$$\epsilon(t) = \frac{1}{16\pi G} \left(\frac{V'(\bar\varphi(t))}{V(\bar\varphi(t))}\right)^2 , \qquad \delta(t) = \frac{1}{16\pi G}\left(\frac{V'^2(\bar\varphi(t))}{V^2(\bar\varphi(t))} - 2\frac{V''(\bar\varphi(t))}{V(\bar\varphi(t))}\right) . \qquad (4.42)$$
Then one has too
$$r(q_R) \equiv \frac{4\lfloor D_{q_R}^0 \rfloor^2}{\lfloor R_{q_R}^0 \rfloor^2} = 16\, \varepsilon(t_{1q_R}) . \qquad (4.43)$$

Now, as we told, for a comparison with the data, what turns out particularly significant are the angular correlations between temperature–temperature, temperature-polarization, polarization-polarization of the CMB fluctuations. In the TT case such correlation is conveniently expressed by the coefficients of its expansion in terms of spherical harmonics
$$C_{TT\, l} = \frac{1}{4\pi} \int d^2\boldsymbol{n} \int d^2\boldsymbol{n}'\ P_l(\boldsymbol{n}\cdot\boldsymbol{n}')\, \Delta T(\boldsymbol{n})\, \Delta T(\boldsymbol{n}') , \qquad (4.44)$$
$\Delta T(\boldsymbol{n})$ e $\Delta T(\boldsymbol{n}')$ being the deviations from the average temperature in the directions $\boldsymbol{n}$ and $\boldsymbol{n}'$ and $P_l(x)$ the Legendre polynomial of order $l$. Somewhat similar expressions in terms of modified spherical harmonics can be given for the correlations involving the polarization.

By assuming for $\lfloor R_q^0 \rfloor^2$ a simply power form of the type (4.38 a) and solving numerically as done in literature [cp- 15, caps. 6,7] the mentioned complicate equations that describe the evolution of the Universe after $t_{2q_R}$, the coefficient s $C_{TT\,l}$, $C_{TE\,l}$ and $C_{EE\,l}$ can be calculated. Then, if appropriate values ar given the to the exponent $n_s$ and to the coefficient $\lfloor N_S \rfloor^2$, one finds that these are very well reproduced in an entire range, like $30 \leq l \leq 2500$. On the basis of refs. [15] (based on WMAP) and [18], we assume

$$n_s = 0.966 \pm 0.003 \,, \quad \lfloor N_S \rfloor^2 = 1.93 \pm 0.12 \cdot 10^{-10} \qquad (4.45)$$

to be considered experimental data.

In turn such values can be reobtained by eqs. (4.40-42) and the numerical resolution of the system (4.10-11), if we give to the parameters jn the potential the values (1.27).

In fact once H(t) has been evaluated, $a(t)$ can be expressed in its term as

$$a(t) = a_I \exp\left\{-\int_t^{t_I} dt' \, H(t')\right\}, \qquad (4.46)$$

where $t_I$ is the time at which the inflation is ended and $a_I$ stays for $a(t_I)$. Then, takimg the logaritms, the equation for $t_{1q_R}$ can be rewritten as

$$\int_{t_{1q_R}}^{t_I} dt' \, H(t') = \ln \frac{H(t_{1q_R})}{q_R/a_I} \,. \qquad (4.47)$$

As apparent from Fig. 1 or 2 we can take $t_I = 5 \times 10^{-12}$ GeV$^{-1}$. Furthermore, since $t_L - t_I$ is small ($\sim 3.5 \times 10^5$ y) with respect to present life of Universe ($t_0 - t_I \sim 14 \times 10^9$ y) and for $t > t_I$ the expansion is dominate by radiation or matter, we can identify $a_I$ with $a_L = a(t_L)$, $t_L$ being the last scattering time and $t_0$ the present time.

Then, let us denote by $z_L$, $r_L$ and $d_A$ the redshift, the radial coordinate and the angular distance of the last scattering surface. If we set $a(t_0) = 1$, current values are

$$a_L = \frac{1}{z_L+1} \cong \frac{1}{1090}, \quad d_A = 12.99 \text{ Mpc} = 2.034 \times 10^{39} \text{ GeV}^{-1} \,,$$

$$r_L = \frac{d_A}{a_L} = 2.217 \times 10^{42} \text{ GeV}^{-1} \qquad (4.48\text{ a})$$

and

$$\frac{q_R}{a_I} \cong \frac{q_R}{a_L} = 3.490 \times 10^{-37} \text{ GeV} \,. \qquad (4.48\text{ b})$$

Note that for properties of the Bessel functions occurring in the expressions of $C_{TT\,l}$, $C_{TE\,l}$ and $C_{EE\,l}$ such quantities are dominated by values $q \sim \frac{l}{r_L}$ for a given $l$. In particular $q_R$ correspond to $l \sim 700$.

From the value (4.48 b), by solving equation (4.47) with a linear interpolation between $t/10^{-12}$ GeV$^{-1} = -1$ and $-2$, we find (Fig. 3 and Appendix))

$$t_{1q_R} = -1{,}48 \times 10^{-12} \text{ GeV}^{-1} \,, \qquad (4.49)$$

then

$$\bar{\varphi}(t_{1q_R}) = 21.46 \times 10^{19} \text{ GeV} \,, \qquad H(t_{1q_R}) = 7{,}75 \times 10^{13} \text{ GeV} \qquad (4.50)$$

and by (4.42-44)

$$\varepsilon(t_{1q_R}) = 0.0054 \,, \qquad \delta(t_{1q_R}) = 0.0012. \qquad (4.51)$$

Finally from (4.40,41) we have

$$n_s = 0.97 \,, \qquad \lfloor N_S \rfloor^2 = 1.90 \times 10^{-10} \,, \qquad (4.52)$$

that are consistent with (4.45). From (4.43) and (4.51), however, we have $r = 0.086$, that, as we told, violates the experimental bound $r < 0.032$ (see ref. [19]).

As told in sec. 1, we can make $r = 0$ and eliminate the above disagreement and all difficulties related to the quantization of the gravitational field if we assume the latter to be a purely classical phenomenon. This can be done in the simplest way in our formalism by setting $\hat{h}_{\rho\sigma} = 0$ in eq. (2.1), and so replacing everywhere the quantum quantity $\hat{\gamma}_{\rho\sigma}$ with the purely c-numeric $\bar{g}_{\rho\sigma}$ and suppressing the term $S_\gamma$ and the integration over $h_{\rho\sigma}$ and $h'_{\rho\sigma}$ in (3.1).

Then all quantities involved in equations (4,17-21) or (4.21,27, 28) and in particular $D_q(t)$ have to be set equal to zero. Of the equations discussed above only eq. (4.23) remains with the right end side replaced by 0. The last circumstance does not modify the expression of $R_q^0$, that was already deduced by neglecting such term up the tine $t_{1q}$. So the conclusions for the scalar fluctuations remain unchanged.

Obviously, as consequence of the classical Einstein equation (1.4), equations of the type (4.21,27, 28) would remain valid as classical equations with the source expressed in terms of the classical energy-momentum $t_{\mu\nu}$ and all the usual treatments in classical General Relativity would continue to hold.

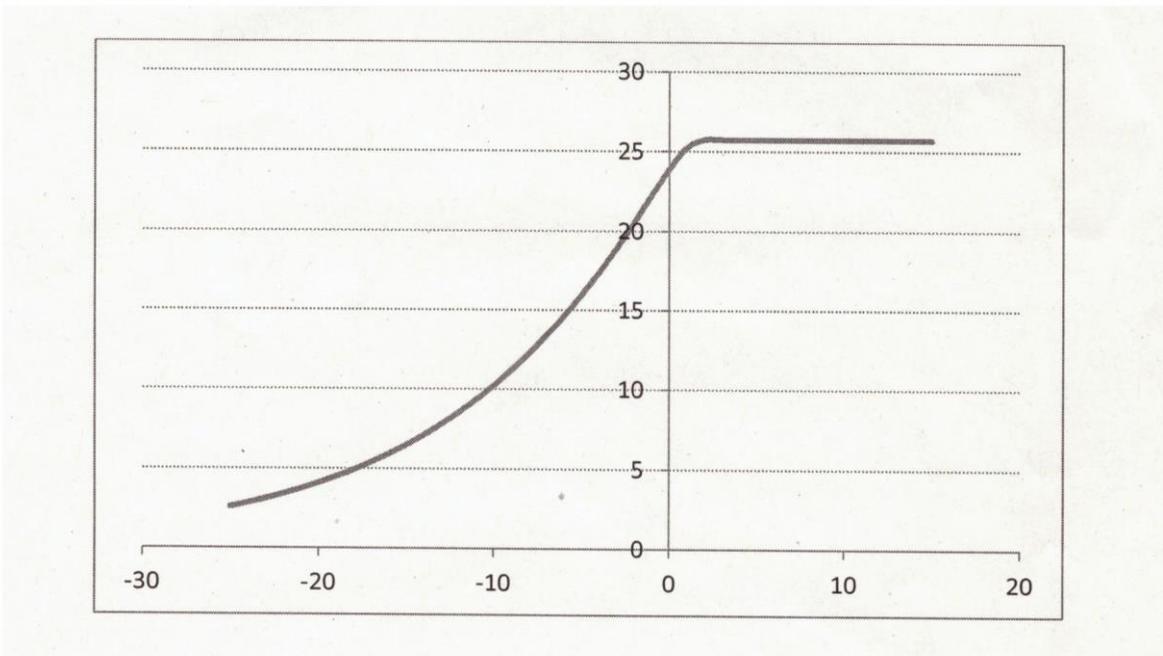

**Fig. 1.** Plot of $\bar{\varphi}(t)/10^{19}$ GeV as a function of $t/10^{-12}$ GeV$^{-1}$, for $\kappa = 8.38 \times 10^{12}$ GeV and $\lambda = 1.05 \times 10^{-15}$ (see (1.27) ). Limit value: $\bar{\varphi} = v = 25{,}7 \times 10^{19}$ GeV.

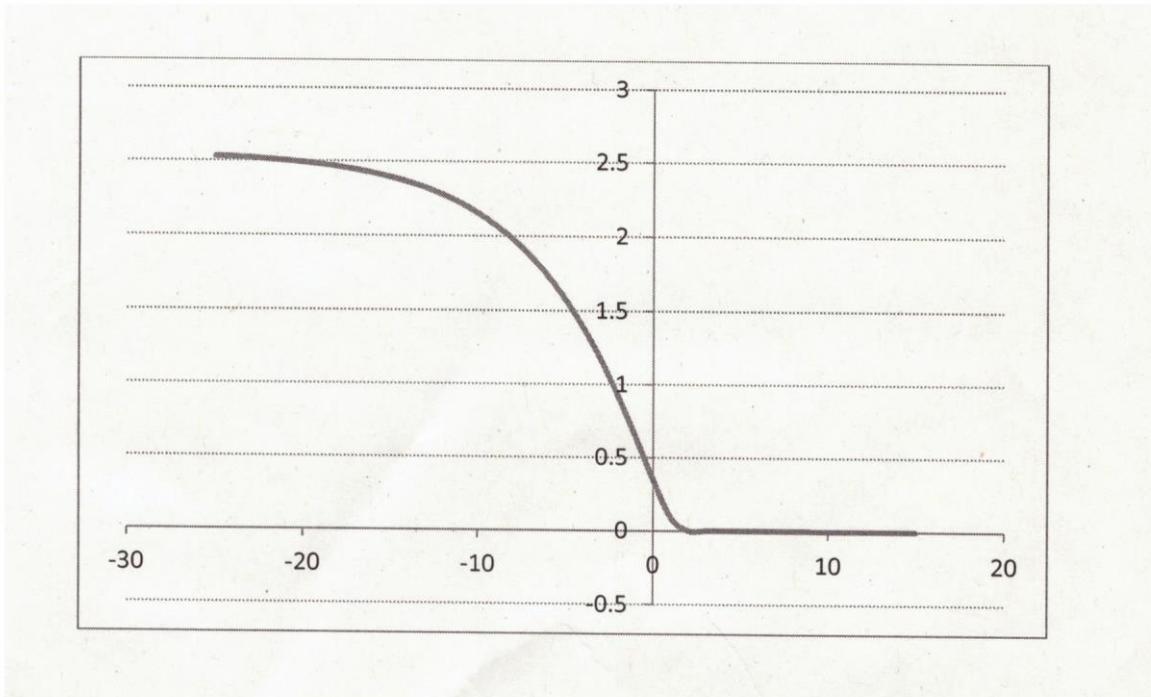

**Fig. 2**. Plot of $H(t)/10^{14}\text{GeV}$ as a function of $t/10^{-12}\text{ GeV}^{-1}$, for the same values of $\kappa$ and $\lambda$. Note $\bar{H} = 2.53 \times 10^{14}\text{ GeV}$ and $t_I \sim 3 \times 10^{-12}\text{ GeV}^{-1}$.

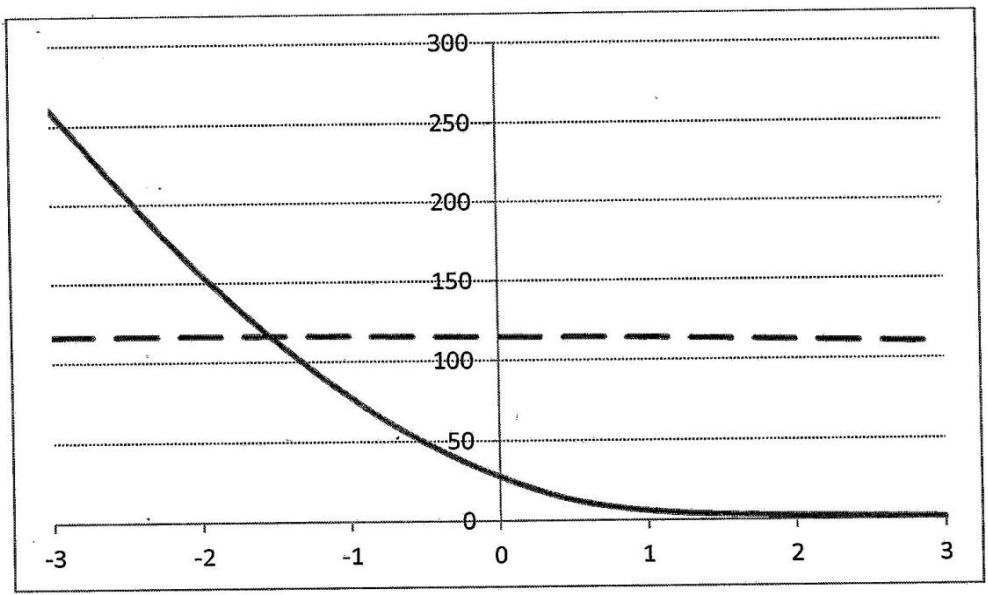

**Fig. 3**. Intersection of the plots:
a) (broken line) $\ln \frac{H(t)\, a_I}{q_R} \equiv 116 + \ln \frac{H(t)}{10^{14}\text{ GeV}}$ ; b) (full line) $\int_t^{t_I} dt'\; H(t')$ ;
 as functions of $t/10^{-12}\text{GeV}^{-1}$.
For interpolation between $t = -2 \times 10^{-12}\text{ GeV}^{-1}$ and $t = -1 \times 10^{-12}\text{ GeV}^{-1}$ :
$t_{1q_R} = -1{,}48 \times 10^{-12}\text{ GeV}^{-1}$, $\bar{\varphi}(t_{1q_R}) = 21.46 \times 10^{19}\text{ GeV}$, $H(t_{1q_R}) = 7{,}75 \times 10^{13}\text{ GeV}$,

## 5. Classical fluctuation

In the preceding section we have explicitly assumed classical fluctuations, as defined in (1.24), to be negligible and presently we have no indication of their occurrence in any other knpwn phenomena. Actually, as we told, we can consistently set the constants appearing in eq, (2.1) equal to zero at the end and consider their introduction simply as a trick to make well defined certain intermediate step. Beside other things this would have the advantage to do not intrude any new undetermined constants in the theory. However we could ask whether we can obtain at least any upper bound for such constants..

To be specific we shall refer to one component of the energy momentum tensor, the energy density $\rho(x) = t_{00}(x)$ and consequently we shall restrict our attention to the constant $\mu$ alone.

From eq. (3,10), at zero order in the quantum gravitation field (or in absence of it) and for a weight function of the form (1.18,19), we have

$$\langle \rho_f^2 \rangle_C = \frac{\Lambda_t \Lambda_s^3 \mu^4}{2\,(2\pi)^2)} ,  \qquad (5.1)$$

by using the convention $a(t_0) = 1$ he quantity $\langle \rho_f^2 \rangle_C$ goes to infinity as $\Lambda_t$ or $\Lambda_t$ increases. However the value of such quantities is fixed by the condition of the experiment and in any case it is limited by the largest energy attainable (of the order $10^4$ GeV today).

Let us introduce the quantity

$$\sigma^2 = \langle \rho_f^2 \rangle_C / \langle \left(\hat{\rho}_f - \langle \rho_f \rangle\right)^2 \rangle_Q \qquad (5.2)$$

that expresses the relative relevance of the classical to the quantum fluctuation.

It can be checked that during the inflation $\sigma^2$ turns out to be extremely small for reasonable values of $\mu$, $\Lambda_t$ and $\Lambda_s$ and so $\langle \rho_f^2 \rangle_C$ irrelevant, as we have supposed. On the contrary the most appropriate field to look for an evidence of a classical variance seems to be the Physics of the Elementary Particles. However, since the standard model of the particles is consistent with the data in some case up to 0.1 %, it is natural to assume $\sigma^2 < 10^{-3}$ at the present energies. To evaluate a typical order of quantum variance in such Physics we find convenient to refer to the decay of a charged particle.

A charged particle is typically observed by the ions it produces in a medium filling the apparatus. To measure the lifetime of an unstable particle, we could count the number of particles of a beam that reach a detector placed at a given distance (*time*) $\bar{t}$, before decaying. Then, if we denote by $\Gamma_Q$ the probability of decay for unity of time according QT ($\tau = 1/\Gamma_Q$ being the lifetime), the probability that a specific particle reaches the required distance can be assumed as $e^{-\Gamma_Q \bar{t}}$. Then, if $\rho_0$ is the density of the medium and the particle actually reaches the point $\bar{t}$, immediately afterwards the weighted energy density in a neighbourhood becomes $\rho_f = \rho_0 + \delta\rho_f$ ($\delta\rho_f$ being the energy deposited as a consequence of the ionization produced). On the contrary if the particle does not reach the point it remains $\rho_f = \rho_0$. So we can write

$$\langle \rho_f \rangle = (\rho_0 + \delta\rho_f)\, e^{-\Gamma_Q \bar{t}} + \rho_0 \left(1 - e^{-\Gamma_Q \bar{t}}\right) = \delta\rho_f\, e^{-\Gamma_Q \bar{t}} + \rho_0 , \qquad (5.3)$$

$$\langle \hat{\rho}_f^2 \rangle_Q = (\rho_0 + \delta\rho_f)^2\, e^{-\Gamma_Q \bar{t}} + \rho_0^2 \left(1 - e^{-\Gamma_Q \bar{t}}\right), \qquad (5.4)$$

and

$$\langle(\hat{\rho}_f - \langle \rho_f \rangle)^2\rangle_Q = (\delta\rho_f)^2 \left(e^{-\Gamma_Q \bar{t}} - e^{-2\Gamma_Q \bar{t}}\right). \tag{5.5}$$

Further, if $\bar{t}$ is of the order of the lifetime $1/\Gamma_Q$, the factor $\left(e^{-\Gamma_Q \bar{t}} - e^{-2\Gamma_Q \bar{t}}\right)$ is of the order $10^{-1}$ and, in particular, if we explicitly take $\bar{t} = 1/\Gamma_Q$, from (5.3) we have

$$\sigma^2 = \frac{e^2}{e-1} \frac{\langle \rho_f^2 \rangle_C}{(\delta\rho_f)^2}. \tag{5.6}$$

Denoting by $b$ the radius of the ion wake produced by the particle, by $\frac{dE}{dx}$ the energy lost for unit length and setting $\frac{1}{\Lambda_t} = \frac{1}{\Lambda_s} = b$, we have $\delta\rho_f = \frac{1}{\pi b^2} \frac{dE}{dx}$ and so

$$\sigma^2 = \frac{e^2}{8(e-2)} \frac{\mu^4}{(\frac{dE}{dx})^2}. \tag{5.8}$$

At the minimum ionization the quantity $\frac{dE}{dx}$ turns out to be independent of the mass of the particle but it is proportional to the density of the medium. For a particle with a unitary elementary charge in air, with a density $\rho_0 = 0.0012 \frac{g}{cm^3}$, we have

$$\frac{dE}{dx} = 2{,}76 \text{ keV/cm} \sim 5 \times 10^{-20} \text{ GeV}^2 \tag{5.9}$$

and so

$$\sigma^2 \sim 10^{38} \mu^4 / \text{GeV}^4. \tag{5.10}$$

Then, under the assumption $\sigma < 10^{-3}$, we obtain

$$\mu < 10^{-11} \text{ GeV} = 0.01 \text{ eV}. \tag{5.12}$$

## 6. Conclusion

Ssummarizing, to make consistent the Copenhagen interpretation of Quantum Theory, we have assumed that a classical or macroscopic state can be assigned to the entire universe, specified in terms of a classical energy momentum tensor, a classical metric, a classical electromagnetic current and other classical currents related to those conserved in elementary particle theory.

Such classical quantities are ideally assumed to have a well determined value at any time and at any points of the space. However their dynamics is supposed to be expressed by a probability distribution for their average values in small space-time domains, which is built in terms of a density operator, the energy momentum tensor and the appropriate current operators in Heisenberg picture. So it is the underling quantum dynamics that drives the macroscopic one.

The important point in which we diverge from the usual text book Quantum Mechanics is that in our perspective any observable occurrence has to correspond to a modification of the classical state and no a priori specific role is assigned to the selfadjoint operators. One important consequence is that the reduction of the quantum state can be simply traced back to a construction of a conditional probability from a joint probability (eqs. (1.26) and (2.8,9)).

The form of the classical variance as apparent e.g. from eq. (5.1) indicates that discrepancies with ordinary QT should emerge at very high energy. This opens the possibility in principle of an experimental test of the theory and of a determination of the constants which appear in eq. (2.1). However a zero value for such constants is also perfectly compatible with the theory. In any case at the presently attainable energies, the standard model of the particles, based on the ordinary QT, turns out to be reasonably satisfatory and only an upper bound of the type (5.12) can be obtained.

For the Universe a variant of the standard ΛCDM is assumed, with $\Omega = 1$, the Big Bang (BB) (understood as a time in the past at which $a(t) \to 0$) shifted at $t = -\infty$, a single inflaton with an Higgs type potential and initial vacuum quantum state. In this way the total energy density is related simply to the constant term in the inflaton potential.

With an appropriate choice of the parameters of the potential, scalar fluctuations in the CBM consistent with the data can be obtained in our as in more conventional models. It remains the difficulty of the lack of tensor fluctuations, which suggests a pure classical character of the gravitational field. As we mentioned this could be consistently obtained within our formalism by suppressing the first term in eq. (1.7) and replacing everywhere the quantum metric tensor $\gamma_{\mu\nu}$ with the tensor $\bar{g}_{\mu\nu}$ defined by eqs. (1.14-16).

The problem of the present acceleration in the expansion of the Universe and of the dark energy can be treated in the standard way. It could be added a small cosmological constant to the constant term in the inflaton potential. Alternatively a new scalar field (quintessence) could be introduced, analogous to the inflaton one but with appropriate values of the potential parameters and for the constant $c$ in the initial condition (4.12).

In the context, it may be interesting to note that the shift to $-\infty$ of the BB is not optiona, but it is imposed by the other hypothesis. In fact let us consider the general Friedman equation

$$\dot{a}^2 + K = \frac{8\pi}{3} G \rho a^2 \ . \tag{6.1}$$

For an energy density $\rho$ equal to a constant, $\bar{\rho}$ its general solution can be written

$$a(t) = \bar{a} \, e^{\bar{H}t} + \frac{K}{4\bar{a} \bar{H}^2} e^{-\bar{H}t} \ , \tag{6.2}$$

where $\bar{a}$ is an arbitrary constant and $\bar{H} = \sqrt{\frac{8\pi}{3} G \bar{\rho}}$. Then, under the assumption that all field vanish at the BB, $a(t)$ must approach the form (6.2). Now for $K = 0$ ($\Omega = 1$) (6.2) reduces to (4.12) and so the BB must be at $-\infty$, as we supposed; for $K > 0$ ($\Omega > 1$), $a(t)$ has no zero and there is no BB, only for $K < 0$ ($\Omega < 1$) we could have a BB at a finite $t$.

Let us conclude with a consideration concerning the basic postulate. The objection could be raised that no sense has to define a probability for the entire Universe, which is a single system. Actually such a criticism would be on the basis of a frequency interpretation of probability, while in a sense our attitude is rather that of a subjective interpretation. In any case, note that in fact our postulate provides correlations among the classical variables in different always finite regions of space-time. So, across the space time similar experiments can be repeated and frequencies predictions tested by experiments.

**Acknowledgments.**

We have to thank our fiends M. Bersanelli, S. Bonometto, L. Lanz , M. Pauri, L. Molinari for useful comments and discussions-

**Appendix**

**Tab. 1**. Numerical resolution of the system (1.15,16) and related quantities for $\kappa = 8.38 \times 10^{12}$ GeV, $\lambda = 1.05 \times 10^{-15}$. The line corresponding to $t = t_{q_R} = -1.48 \times \mathbf{10^{-12}}$ GeV$^{-1}$ is obtained by linear interpolation.

| $t/10^{-12}$ GeV$^{-1}$ | $\bar{\varphi}(t)/10^{19}$ GeV | $H(t)/10^{14}$ GeV | $\int_t^{t_{I_I}} dt'\, H(t')$ | $\ln \dfrac{H(t)\, a_I}{q_R}$ |
|---|---|---|---|---|
| -25 | 2.66 | 2.53 | 5083 | |
| -23 | 3.13 | 2.52 | 4473 | |
| -21 | 3.76 | 2.50 | 3970 | |
| -19 | 4.51 | 2.48 | 3173 | |
| -17 | 5.41 | 2.44 | 2980 | |
| -15 | 6.50 | 2.39 | 2394 | |
| -13 | 7.79 | 2.32 | 2323 | |
| -11 | 9.34 | 2.22 | 1570 | |
| -9 | 11.19 | 2.07 | 1140 | 116.9 |
| -7 | 13.39 | 1.87 | 751 | 116.8 |
| -5 | 15.97 | 1.57 | 406 | 116.7 |
| -3 | 18.96 | 1.17 | 260 | 116.2 |
| -2 | 20.59 | 0.920 | 156 | 116.1 |
| -1.48 | 21.46 | 0.775 | 115.4 | 115.9 |
| -1 | 22.27 | 0.642 | 78 | 115.8 |
| 0 | 23.91 | 0.349 | 28 | 114.6 |
| 1 | 25.26 | 0.093 | 6 | 113.8 |
| 3 | 25.74 | 0.002 | 1 | 110.1 |
| 6 | 25.73 | 0.001 | 1 | 99.1 |
| 9 | 25.74 | 0.001 | 0 | |
| 12 | 25.73 | 0.000 | 0 | |
| 15 | 2574 | 0.000 | | |

$x$